# Mapping Topics and Topic Bursts in PNAS

Ketan Mane & Katy Börner
*Indiana University, SLIS*
*10th Street and Jordan Avenue*
*Bloomington, IN 47405 USA*
*Email: [1] @ indiana.edu*

**Scientific research is highly dynamic. New areas of science continually evolve; others gain or lose importance, merge or split. Due to the steady increase in the number of scientific publications it is hard to keep an overview of the structure and dynamic development of one's own field of science, much less all scientific domains. However, knowledge of 'hot' topics, emergent research frontiers, or change of focus in certain areas is a critical component of resource allocation decisions in research labs, governmental institutions, and corporations. This paper demonstrates the utilization of Kleinberg's burst detection algorithm, co-word occurrence analysis, and graph layout techniques to generate maps that support the identification of major research topics and trends. The approach was applied to analyze and map the complete set of papers published in the Proceedings of the National Academy of Sciences (PNAS) in the years 1982-2001. Six domain experts examined and commented on the resulting maps in an attempt to reconstruct the evolution of major research areas covered by PNAS.**

### Introduction

Maps depicting the structure and evolution of scientific fields are also called *Knowledge Domain Visualizations* (KDVs) [2]. They are a special kind of *Information Visualization* [3] that exploit powerful human vision and spatial cognition to help humans mentally organize and electronically access and manage large, complex information spaces. Unlike scientific visualizations, KDVs are created from data that have no spatial reference, such as sets of publications, patents, or grants.

KDVs use sophisticated data analysis and visualization techniques to objectively identify major research areas, experts, institutions, grants, publications, journals, etc., in a domain of interest. They can be used to gain an overview of a knowledge domain; its homogeneity, import-export factors, and relative speed; to track the emergence and evolution of topics; or to help identify the most productive as well as new research areas. Benefits of KDVs include reducing visual search time, revealing hidden relations, displaying data sets from several perspectives simultaneously, facilitating hypothesis formulation, serving as effective means of communication, and prompting users to think in new ways about document data. Today, KDVs are typically generated semi-automatically from rather small, static data sets, and for a specific knowledge domain and information need.

This paper presents a novel way to generate co-word association maps of major topics based on highly frequent words and words with a sudden increase in their usage, a phenomenon called 'burst' [4]. A large-scale data set comprising the complete set of 47,073 papers published in the Proceedings of the National Academy of Sciences (PNAS) in the years 1982-2001 is used for demonstration. We describe the identification of highly frequent and bursty words, the analysis of the most important correlations among those words, and the generation and interpretation of a two-dimensional layout showing major research topics and their dynamics.

### Tracking the Evolution of Major Topics

To identify major words or topics covered in PNAS, we first selected the top 10% of the most highly cited documents for each of the 20 years. This is common practice, as papers with few citations are assumed to have less impact and most algorithms simply cannot handle very large amounts of data. The least cited paper in this set received 14 citations.

The next step is the identification of major sources for potential topic words. Biologists are well aware that titles and keywords are used for indexing. Hence, they tend *not* to use words that occur in the title as keywords and vice versa. Therefore, paper titles and keywords were selected for the subsequent analysis. Two types of keywords exist: ISI keywords, which come from the author, publisher, or titles of cited papers and Medline's controlled vocabulary, also called MeSH terms. No ISI keywords are available for papers published prior to 1991. MeSH terms have been joined to ISI records using the procedure by Boyack [5]. Papers without titles were excluded from the analysis, resulting in 4,699 papers. From those papers a total of 34,299 unique potential topic words were extracted.

In order to determine the trends of word usage over time, the top ten most frequent *and* meaningful words were then selected in collaboration with domain experts. Those words in the order of decreasing frequency are *human, animal, mice, molecular sequence data, genes, expression, RNA, DNA, cell line,* and *cloning*. Excluded from the top frequency list were *support, U.S. gov't, non-U.S. Gov't, P.H.S., receptors, cells, rats, amino acid sequence, base sequence*, and *cultured*.

Figure 1 shows the frequency count for all 10 words for the 20-year time period. Clearly visible is the introduction of new terms as well as the increase or decrease in the usage of certain words and the influence of biotechnology events [6]. For example, *human* studies are steadily increasing and are bursting at the start of the Human Genome Project in 1988. Research on *animal* and *mice* shows a similar trend. Research on *genes*, *DNA*, and *RNA* is strongly coupled and shows similar upward trends.

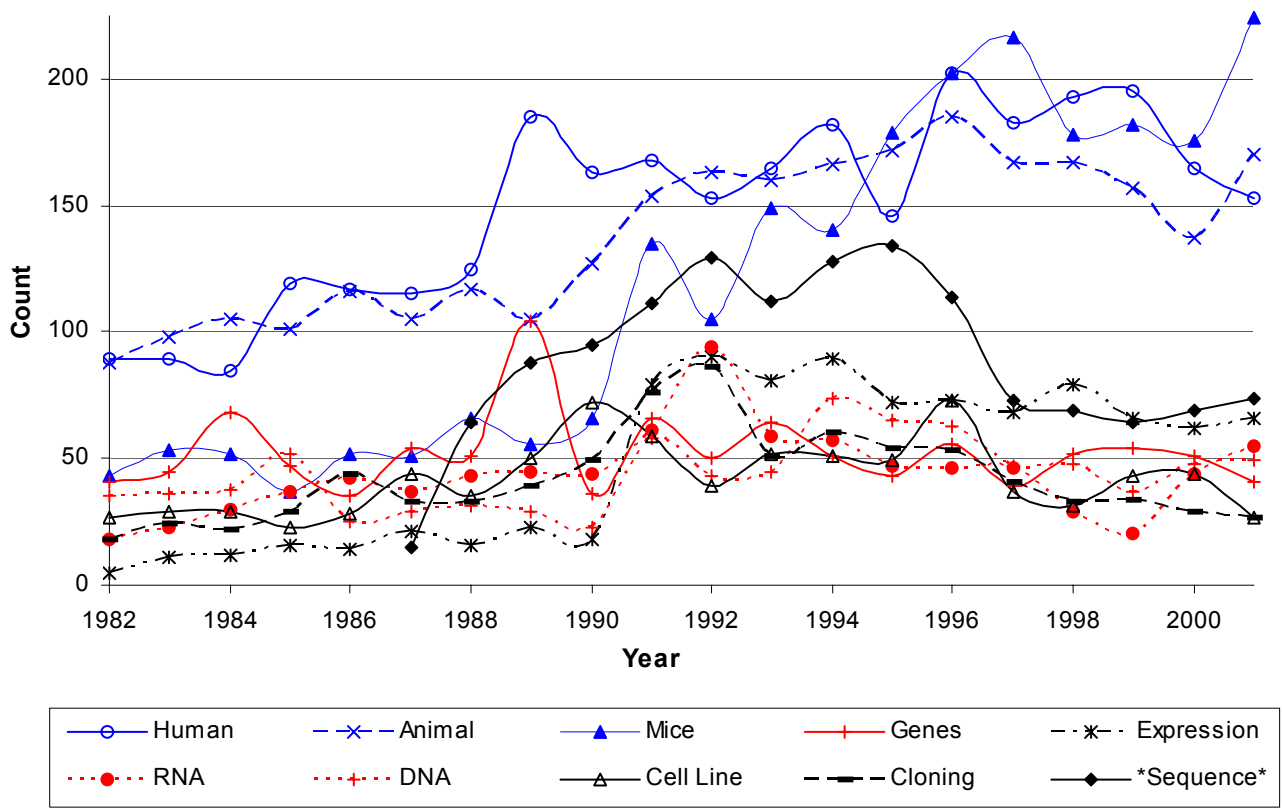

**Figure 1** Frequency count for most frequently used words in the top 10% of most highly cited PNAS publications from 1982 – 2001. The color version of the figure is available at: http://ella.slis.indiana.edu/~katy/gallery/03pnas-fig1.jpg

As exemplified in Figure 1, several research trends have occurred in parallel to word frequencies. For example, polymerase chain reaction (PCR) technology, conceived in 1983 and combined with reverse transcription, was fully developed by 1986 and commonly utilized for gene *expression* experiments. *Cell-line* research was carried out during the 1982-1991 time period. It declined in later years as focus shifted from research to application development. Molecular *sequence* data research received a boost with the sanction of funds for the Human Genome Project. Its importance decreased after 1994 as it becomes routine in genetic studies. In 1991, work on the *expression* rate of DNA to protein conversion (rather than just the transcription process) ignited. As more data on molecular *sequence*, *expression* rate, etc. became available, research towards developing molecular models and protein modeling became relevant. *Cloning* research had its first breakthrough result with the successful cloning of a sheep in 1997. Molecular cloning approaches were quickly adopted in the subsequent years.

In addition to word frequency, the dynamics of topic usage were also examined. Kleinberg's burst detection algorithm [4] was applied to identify topics that experience a sudden increase in their usage, also called burst. The burst detection algorithm provides a formal model for the robust and efficient identification of word bursts. Using a finite state automaton, bursts in streams of words correspond to state transitions. The algorithm outputs the start and end time of a burst as well as its strengths for each word. Some words in the PNAS data set experience multiple bursts. In the top 10 % of the most highly cited PNAS publications there were 1027 unique words, of which 991 had at least one burst and 34 of which had two bursts. Exactly two words: 'comparative study' and 'dna primers' bursted on three occasions but are not among the highly frequent words.

**Mapping the Co-Occurrence Space of Topics and Topics Bursts**
After analyzing the frequency occurrence of all 34,299 unique words for each of the 20 years as well as their burstyness, we next chose to identify and map the relationships among major topics.

Co-word occurrence analysis is a content analysis technique that can be used to identify the strength of associations between words based on their co-occurrence in the same document [7]. Words that appear together often will have a strength closer to 1, and words that never appear together a strength of 0. While co-word spaces are typically generated based only on highly frequent words, the work presented here is unique as it also accounts for word burstyness.

To begin, we computed the intersection of the highest frequency and most highly bursting word sets and selected the first 50 for further analysis. Interestingly, there was a rather low correlation among the frequency of words and their burstyness. In the particular example discussed here, it took 742 most frequent words and 874 most bursty words to get an intersection of 50 words.

The co-word analysis was conducted for those 50 words and the set of 4,699 documents. The resulting co-occurrence frequency matrix was normalized using Salton's cosine coefficient [8] where each word pair co-occurrence is defined as the ratio of their co-occurrences and the product of the square root of the respective word occurrences within the document set. Interestingly, the original, non-normalized co-occurrence matrix resulted in more meaningful maps as judged by domain experts and will be used in the subsequent discussion.

The non-normalized co-occurrence matrix has 1,082 non-zero entries characterizing the complex co-occurrence relationships among the 50 words. In order to reduce this number to the most meaningful relationships, the pathfinder network scaling algorithm (PFNet) [9] was applied. The PFNet algorithm relies on triangle inequality to eliminate unnecessary links. Given two paths (sequence of links) in a network that connects two nodes, the path that has a greater weight as defined via the Minkowski metric is preserved. It is assumed that a path with a greater weight better captures the interrelationship between two nodes and that alternative paths with less weight are redundant or even counter-intuitive and should be pruned from the network. Two parameters, *r* and *q* influence the topology of a pathfinder network. The r-parameter specifies the weight of a path based on the Minkowski metric. The q-parameter defines the number of links in alternative paths (i.e., the length of a path) up to which the triangle inequality must be maintained. A network of N nodes can have a maximum path length of q=N-1. With q=N-1 the triangle inequality is maintained throughout the entire network. A detailed explanation of PFNet as well as another application of it is reported by Chen [1].

Running PFNet with q=N-1=49 and infinite *r* results in 62 non-zero entries and, hence, a very sparsely connected network of 50 topic nodes and 62 edges. The many lattice-like subgraphs in this network reveal only part of the complex relationships among the major topics. Based on expert feedback, we selected the network with 80 edges that was generated using the parameter values r=6 and q=49.

Subsequently, the topic word co-occurrence network was laid out in two dimensions for visual examination. The layout depicted Figure 2 was generated using the Fruchterman-Reingold 2D graph layout algorithms [10], a more efficient version of the original spring embedding algorithm developed by Eades [11]. Each node in the network represents one of the 50 highly frequent and bursting words. Note that the words *human, animal, mice*, and *DNA* discussed in Figure 1 do not burst and are hence not included in Figure 2. The size of the node circle corresponds to the maximum burst level this word achieved. Color-coding is used to denote the years in which the word was used most often as well as the year of the maximum burst. Five time durations and respective colors were used: 1982-85 in green, 1986-89 in yellow, 1990-93 in orange, 1994-97 in red, and 1998-01 in black. The year of the maximum frequency and the starting year of the first burst of this word were decoded by circle border colors and inner circle area colors respectively. For example, the word *molecular sequence data*, represented by a circle of rather large size with an orange inner area and a red ring, showed the highest, rather large burst between 1990-93 and had a high frequency of usage in the later years 1994-97.

Edge thickness is proportional to the number of word co-occurrences. *Protein* and models or cells and growth are co-occurring frequently in the selected publication data set.

The evaluation and interpretation of the resulting map is rather difficult as there are very few people who are familiar with the diverse research results reported in the original data set (4,699 highly cited PNAS papers published over a 20-year time span). However, the graph visualization shown in Figure 2 was examined by six biologists and their interpretation of the map is summarized below.

Over the last 20 years, biological research has experienced enormous growth and also diversification. The graph shown in Figure 2 semantically interrelates and chronologically links diverse fields of biological research. Four major areas can be identified that are interlinked with the middle, oldest area of research. The top left subnetwork is related to expression profiling and genomics research. Top right topics deal with protein research, right bottom are linked to cancer research, and the ones in the left bottom relate to molecular sequence studies.

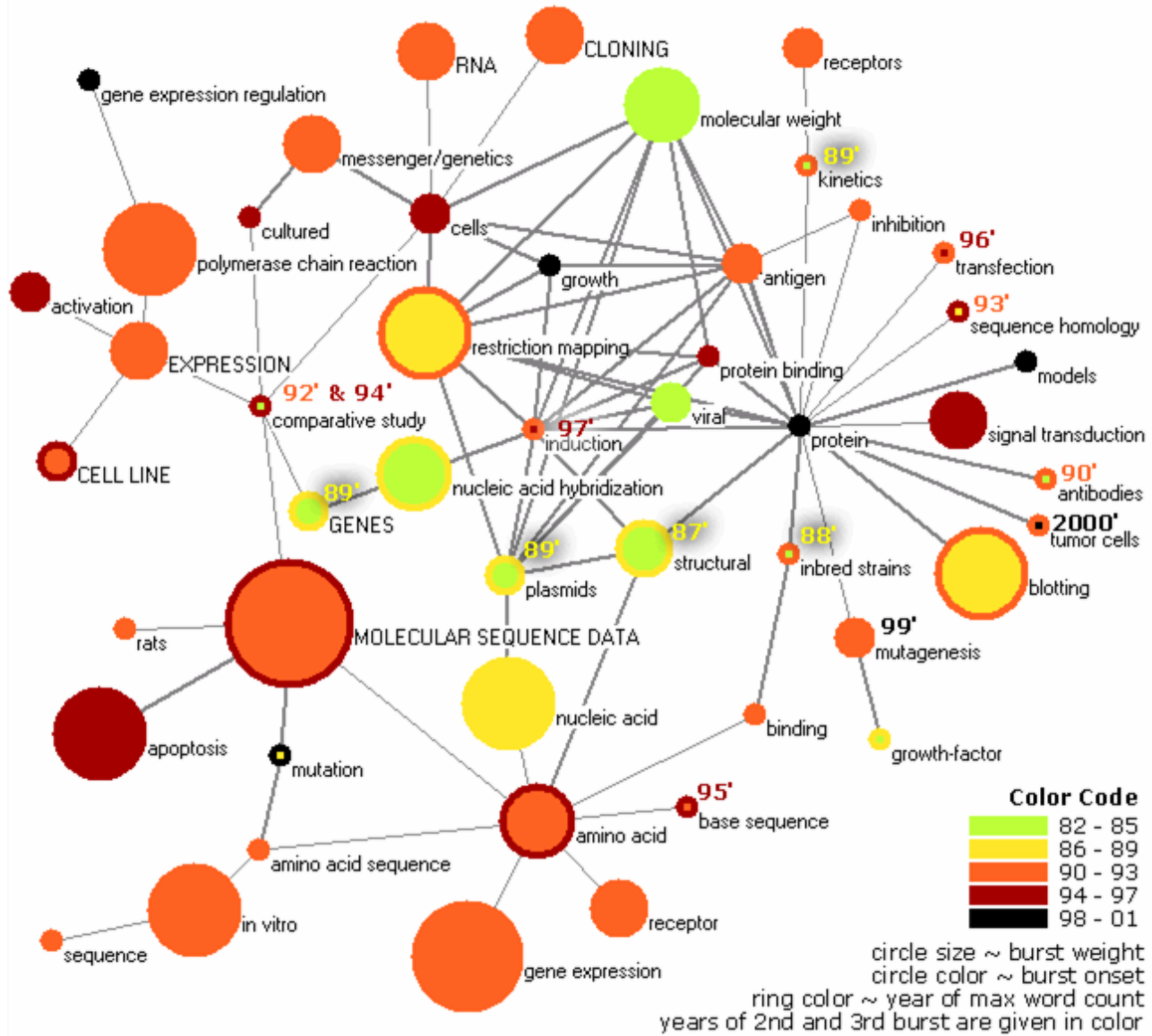

**Figure 2** Co-word space of the top 50 highly frequent and bursty words used in the top 10% most highly cited PNAS publications in 1982-2001. The color version of the figure is available at: http://ella.slis.indiana.edu/~katy/gallery/03pnas-fig2.jpg

In the early 1980's the primary research foci were structural properties of biological entities such as cells, genes, etc. This was followed by a phase of research in kinetics and the study of the mutation behavior of genes. Toward the early 90's, in conjunction with the start of the human genome project, the research paradigm shifted toward sequence data studies. During this time period, molecular sequence data - amino acid sequences associated with the genome project - rose to prominence. Major funding via the human genome project also brought together several interconnected research areas primarily dealing with cloning, PCR, and gene expression depicting cloning studies. These experiments are an extension of prior studies on plasmids, genes, and nucleic acid hybridization. In later years, research concentrated on apoptosis, signal transduction, activation, and cells which are linked by cell signaling pathways for programmed cell death, a key area of cancer research. The increase in computing power facilitated extensive research in modeling research, leading in turn to an increased understanding about the folding patterns of proteins. In 2000, the human genome sequence was completed and investigations now concentrate on protein research.

**Summary**
The paper demonstrated an objective, computational approach to analyze the structure and evolution of a research domain. To our knowledge, this is the very first attempt to map the co-word space of highly frequent and bursty words. The resulting visualization depicts 50 major topics and topic bursts in PNAS and their evolution over a 20-year time frame.

Problems of dimensionality reduction for generating plots of high dimensional data sets were tackled using threshold values to select a representative document and unique word set as well as the application of pathfinder network scaling to capture major associations among words.

The resulting visualizations were examined and interpreted by a number of domain experts, demonstrating their readability and practical value for the identification of topics, major trends, and research frontiers as well as hinting at their value as a knowledge management tool for researchers, companies, funding agencies, and society.

**Acknowledgements**
We would like to thank Anne Prieto, Don G. Gilbert, Sun Kim, Keith G. Ngolley, Kranthi Varala, and Claire Nisonger for insightful comments on the interpretation of the generated maps and Margaret Swan for proof reading this paper. The KNOT tools for Pathfinder Network Analysis[1] and Pajek [12] were used in the presented analysis.

The PNAS data was extracted from Science Citation Index Expanded – the Institute for Scientific Information®, Inc. (ISI®), Philadelphia, Pennsylvania, USA: © Copyright Institute for Scientific Information®, Inc. (ISI®). All rights reserved. No portion of this data set may be reproduced or transmitted in any form or by any means without prior written permission of the publisher.

This work is supported by a National Science Foundation CAREER Grant under IIS-0238261 and NSF grant DUE-0333623.

---

[1] http://interlinkinc.net/